\documentclass[journal]{IEEEtran}
\usepackage{tabularx,booktabs}
\usepackage{multirow}
\usepackage{caption}
\usepackage{subcaption}
\usepackage{amsmath}
\usepackage{hyperref}
\ifCLASSINFOpdf
   \usepackage[pdftex]{graphicx}
   \DeclareGraphicsExtensions{.pdf}
\else
\fi

\begin{document}

\title{A fully recurrent feature extraction for single channel speech enhancement}

\author{Muhammed~PV~Shifas,
        Santelli~Claudio, Vassilis~Tsiaras,
        Yannis~Stylianou,~\IEEEmembership{Fellow ~IEEE}
        \vspace{-0.4cm}
\thanks{This work was funded by the E.U. Horizon2020 Grant Agreement 675324, Marie Sklodowska-Curie Innovative Training Network, ENRICH.}
\thanks{Muhammed PV Shifas, Vassilis Tsiaras and Yannis Stylianou are with the Speech Signal Processing Laboratory, Department of Computer Science, University of Crete, Greece (e-mail: shifaspv@csd.uoc.gr).}.
\thanks{Santelli Claudio is associated with the Sonova AG, Staefa, Switzerland.}}


\maketitle
\begin{abstract}
Convolutional neural network (CNN) modules are widely being used to build high-end speech enhancement neural models. However, the feature extraction power of vanilla CNN modules has been limited by the dimensionality constraint of the convolution kernels that are integrated -- thereby, they have limitations to adequately model the noise context information at the feature extraction stage. To this end, adding recurrency factor into the feature extracting CNN layers, we introduce a robust context-aware feature extraction strategy for single-channel speech enhancement. As shown, adding recurrency results in capturing the local statistics of noise attributes at the extracted features level and thus, the suggested model is effective in differentiating speech cues even at very noisy conditions. When evaluated against enhancement models using vanilla CNN modules, in unseen noise conditions, the suggested model with recurrency in the feature extraction layers has produced a segmental SNR (SSNR) gain of up to 1.5 dB, an improvement of 0.4 in subjective quality in the Mean Opinion Score scale, while the parameters to be optimized are reduced by 25\%.

\end{abstract}

\begin{IEEEkeywords}
Speech enhancement, deep neural network, recurrent features extraction.
\end{IEEEkeywords}

\IEEEpeerreviewmaketitle

\section{Introduction}
%
%
%
%

\IEEEPARstart{s}{peech} enhancement is concerned with improving the intelligibility and/or overall perceptual quality of speech that has been degraded by additive noise. With the increased use of communication devices in noisy environments, the need for robust enhancement strategies is of paramount importance.
Classical speech enhancement techniques deal with the problem of enhancing speech signals that have been degraded by background quasi-stationary noise.
These methods enhance speech by either subtracting the noise magnitude spectrum from the noisy speech spectrum~\cite{boll1979suppression} or by modeling the noise distribution with first and second-order statistics~\cite{ ephraim1992statistical}.
However, the spectral subtraction produces distortions at very low SNRs~\cite{evans2006assessment}.  In addition, the first and second order statistics are not sufficient to separate non-stationary noise from the speech and the intelligibility of the processed noisy speech tend to diminish.

Neural networks (NNs) have attracted attention for the speech enhancement task because they are capable to learn high order statistical information and thus, being able to represent the complex mapping function from noisy to clean speech~\cite{xu2014experimental, lu2013speech}. 
The first applications of NNs in speech enhancements considered fully connected neural networks (FNNs)~\cite{xu2014experimental, lu2013speech}.
FNNs predict an output frame from the corresponding input frame or from a small window of frames around it. In tasks that require long receptive fields, such as separating a target speaker from babble noise, their performance drop~\cite{tan2018convolutional}. Alternative architectures that are capable to model time dependencies efficiently include convolutional neural network (CNN) layers and/or recurrent neural network (RNN) layers. A CNN layer captures the local dependencies and a network of CNN layers can capture longer dependencies. As a result CNNs perform better than FNNs~\cite{park2016fully}. Also, CNNs are more memory efficient than FNNs due to their weight sharing property.  
However, as the depth of the CNNs grows, the number of their parameters also grows and this limits their applicability in low end and embedded devices.
On the other hand, RNNs are capable to model long dependencies with only one or few stacking layers.
Weninger et al.~\cite{weninger2015speech} used LSTMs recurrent neural networks to pre-process and clean speech before using it for noise-robust automatic speech recognition and they reported state of the art word error rate. Networks that combine CNN and RNN layers were also considered recently ~\cite{zhao2018convolutional}, \cite{naithani2017low}. In these networks the CNN layers specialize in feature extraction and the RNN layers in modeling the longer dependencies.

In this work we employ a recurrent cell, called gruCNN, which combines the feature extraction ability of CNN with the long-term memory of GRU cell~\cite{cho2014learning}. The gruCNN cell was first proposed by Hartmann~\cite{hartmann2018seeing} in order to make machine vision robust under imperfect lighting conditions and noisy environments.
The gruCNN use recurrent connections within the CNN’s convolution layers and can learn to integrate information over time. This feature is particularly helpful when there is low signal quality such as low-quality video, taken at night with poor lighting conditions and with motion distortions and with occluded objects~\cite{hartmann2018seeing}. The architecture of gruCNN cell, which was designed to integrate images over time, fits well with the speech enhancement task since in the new task the input is also a sequence of two dimensional spectrograms. This architecture can be seen as an extension of the ConvLSTM-SE model presented in \cite{pvtowards}, where the feature integration over time was done with a long short-term memory cell type architecture with input, forget and output gates. Here, the suggested enhancement model (gruCNN\_FC-SE) utilizes gruCNN cells to learn to extract features that are maximally relevant in every temporal context. The suggested model is robust in modelling speech recursion with minimal parameters.
When trained and evaluated on a multi-speaker data set, under different unseen noise conditions, gruCNN\_FC–SE model provides promising results over the traditional networks. The speech intelligibility is improved, in the segmental SNR scale, up to 1.5 dB, across different SNR levels. At the same time, the number of parameters is reduced by 25\% compared to the traditional recurrent model.

The rest of this paper is structured as follows. In Section~\ref{Sec-2}, we discuss the suggested feature extraction strategy, and the gruCNN\_FC--SE enhancement model. The model evaluation procedure is in Section~\ref{Sec.3}. Section~\ref{Sec.4}, includes the results and discussion on the observations. The paper is concluded in Section~\ref{conclusion}.

\section{Recurrent Feature Extraction Technique}
\label{Sec-2}
The problem of speech enhancement is framed on the manually extracted feature (spectral) domain of speech, for the higher computational complexity of temporal models. Since speech is highly auto-regressive in nature, the speech samples generation should be modelled statistically. Let $X_{k}$ be the slice of $k^{th}$ frequency bin values over time, from the noisy input spectrum $X$, such that $X_{k}$ = [$x_{1}, ...., x_{T-1},  x_{T}$]; where $T $ is the total number of frames considered. Then, the probability of $X_{k}$ to happen can be expressed as     
\begin{equation}
   p(X_{k})= p(x_{1}, ....,  x_{T-1}, x_{T})
   \label{eqn1}
\end{equation}
with the product rule of probability, the joint distribution can be redefined as the product of individual probabilities:
\begin{equation}
   p( X_{k})= \prod_{t=1}^{T} p(x_{t} / x_{t-1}, ..., x_{t-T})
   \label{eqn3}
\end{equation}

Preserving this statistical structure is essential when designing speech enhancement models to ensure the auto-regressive nature of predictions. Moreover, the quality of enhancement will be determined by how accurately this dependency is being modelled. Though there may have been some inter-bin dependencies between the spectral bins within a frame, as k varies from 1 to K (the final bin), present modelling has not considered that for it may be trivial compared to the temporal dependency. With this decomposition, only the past dependencies are considered for the model to be causal.

In conventional speech enhancement neural models~\cite{zhao2018convolutional}\cite{naithani2017low} the temporal recurrency of speech was modelled by fully connected recurrent neural network (FC-RNN) layers, like LSTM~\cite{hochreiter1997long}, employed towards the end of model architecture. Therefore, the front-end feature extraction with CNN layers and the recurrency modelling  with FC-RNN operate independently. Such modeling, without counting recurrency factor at the feature extraction level, leads to the lack of qualitative features at front-end. Further, due to the inherent fully connected nature of FC-RNN, the bin-wise recurrrency factor described in~(\ref{eqn3}) has been ignored. 

In this paper, a new feature extraction strategy utilizing the local recurrency of speech is suggested. The feature extraction layers are designed to model the local recursion over time with kernels of fixed dimension that trace the local statistics of previous frame to be integrated into the current feature estimation. 
At a given frame index $t$, the new feature extraction layer (gruCNN) has inputs the previous layer output $X_{t}$ -- which is the noisy speech spectrum at the beginning layer -- and the feature status of the previous frame ($H_{t-1}$). This is being processed through the nonlinear transformations in~(\ref{eqn4}) -- (\ref{eqn7}) to get the feature representation of the current frame ($H_{t}$). Whereby, the feature map $H_{t}$ encodes information from the current frame together with the past context. 

\begin {equation}
Z_{t}=\sigma (W_{zh}*H_{t-1}+W_{zx}*X_{t})
\label{eqn4}
\end{equation}
\begin {equation}
R_{t}=\sigma (W_{rh}*H_{t-1}+W_{rx}*X_{t})
\label{eqn5}
\end{equation}
\begin {equation}
\hat{H_{t}}=\tanh (W_{hh}*(R_{t} \odot H_{t-1})+W_{hx}*X_{t})
\label{eqn6}
\end{equation}
\begin {equation}
H_{t}=Z_{t} \odot H_{t-1}+(1-Z_{t}) \odot \hat{H_{t}}
\label{eqn7}
\end{equation}
where the operations $*$ and $\odot$ indicate convolution and element-wise matrix multiplication, respectively. The capitalized variables highlight the fact that they are matrices of dimension $[K \times C]$ at every frame instant, where $K$ and $C$ are the dimension of frequency and channel axis, respectively. 
While training in this setting, the network will learn the optimal kernels ($W_{zh}$, $W_{zx}$, $W_{rh}$, $W_{rx}$, $W_{hh}$ and $W_{hx}$) that maximize the local bins recurrency, whereby ensure the best features at the layers. It is worth to note that unlike fully connected RNN cells~\cite{cho2014learning}, \cite{hochreiter1997long} that use matrix operation to model the long-term context, the gruCNN has kernel coefficients that are shared, which in turn reduces the parameter complexity.

\begin{figure}[h]
  \centering
  \includegraphics[width=0.45\textwidth, height=12.2cm]{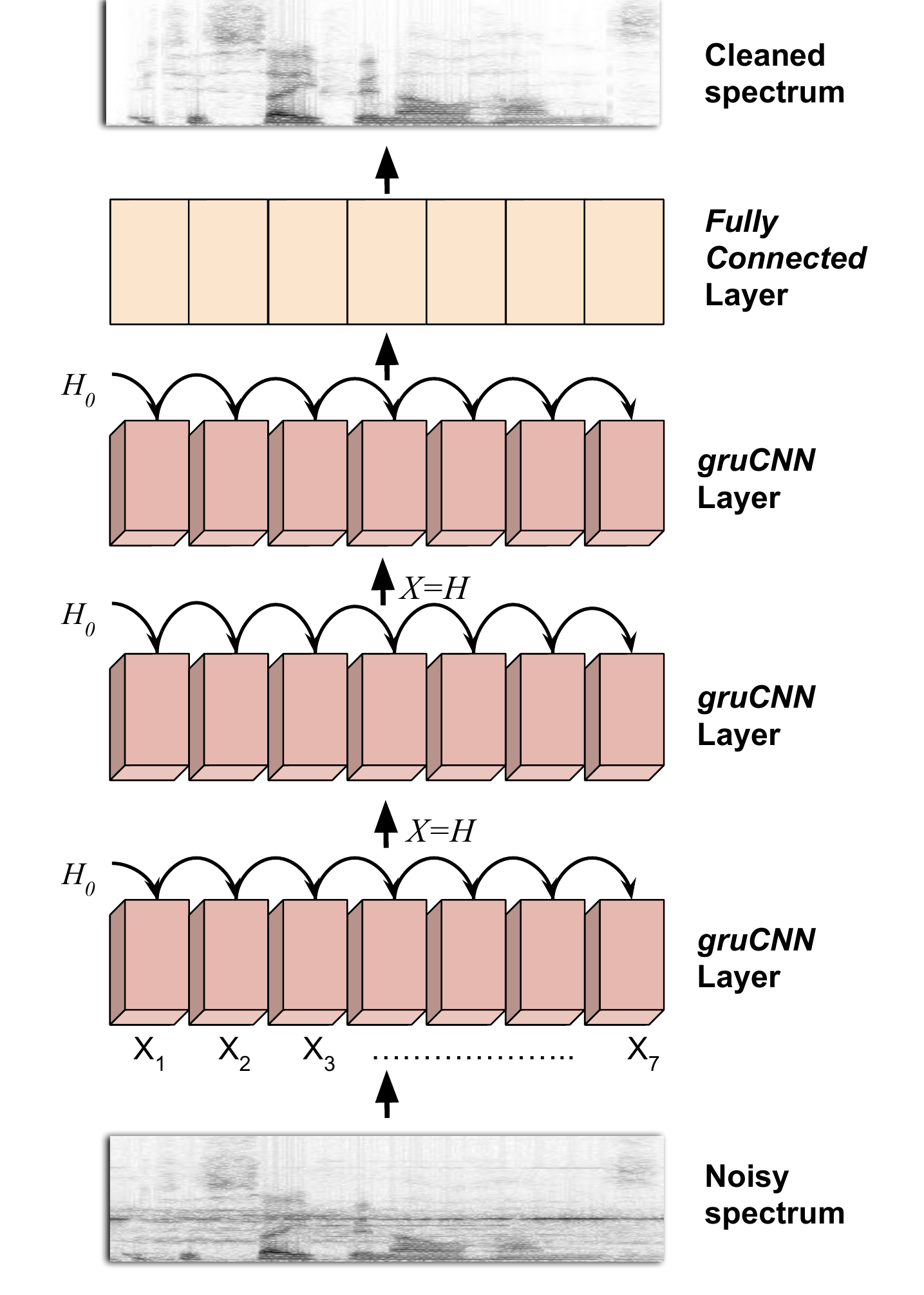}
  \caption{The recurrent feature extraction of $gru$CNN\_FC--SE model. Where $H_{0}$ was set to zero valued tensor at each layer.}
  \label{gruCNN}
\end{figure} 

By layering a set of gruCNN modules one after another, the gruCNN\_FC--SE network has the final structure shown in Fig.~\ref{gruCNN}. At the end of model architecture, it is a time distributed fully connected layer which regresses the recurrently extracted features into the enhanced spectral bins. These predictions are combined with the noisy phase information to reconstruct back the enhanced speech samples.

\section{Evaluation Procedure}
\label{Sec.3}
As the primary focus is on evaluating the efficacy of suggested recurrent feature extraction strategy over the conventional CNN architecture, the comparing models should have had the same structural setting. To this purpose, a model without any recurrent connections in the feature extracting CNN layers is considered (CNN\_FC--SE). Since it does not incorporate any form of temporal recurrency at all in its modeling, the architecture is similar to Fig.~\ref{gruCNN}, but without the recurrent connections. Secondly, to quantify the benefits of recurrency modelled precisely at the feature extraction stage, a model rather having the front-end CNN layers followed by the standard fully connected LSTM cell~\cite{gers1999learning} (CNN\_LSTM--SE) is implemented. The LSTM cell was selected instead of GRU for they have shown better enhancement, as have been reported in the past studies~\cite{zhao2018convolutional}\cite{naithani2017low}. 
 
All the models considered have six convolutional layers (recurrent/casual) followed by the final fully connected (recurrent/casual) layer. The convolutional kernels of each layer are set to be of [3 $\times$ 3] size. The filter size was selected to be of basic for swiftly isolate the performance gain by different models. Each layer of the models has had channel depth of 256 with Parametric ReLU (PReLU) activation. Further details about the individual layers are highlighted in TABLE~\ref{tab:Table-1}, for an input tensor of shape $[1, 161,128,1]$.

\newcommand{\ra}[1]{\renewcommand{\arraystretch}{#1}}
\begin{table}[th]
\caption{Layer-wise description of different models}
 \label{tab:Table-1}
 \centering{
 \ra{1.3}
\scalebox{0.80}{
\begin{tabular}{c c c c c c} 
     \toprule
 {\bf Layer} & {\bf CNN\_FC-SE} & {\bf CNN\_LSTM-SE} & {\bf gruCNN\_FC-SE} & { \bf Output shape} \\ 
      \toprule
 1 &[$3\times3$] CNN  & [$3\times3$] CNN& [$3\times3$] gruCNN & [1, 161, 128, 256] \\ 
  2 & [$3\times3$] CNN & [$3\times3$] CNN& [$3\times3$] gruCNN & [1, 161, 128, 256] \\ 
 3& [$2\times1$] Maxpool &[$2\times1$] Maxpool &[$2\times1$] Maxpool & [1, 81, 128, 256]\\
 4 & [$3\times3$] CNN & [$3\times3$] CNN& [$3\times3$] gruCNN & [1, 81, 128, 256]\\ 
 5 & [$3\times3$] CNN & [$3\times3$] CNN& [$3\times3$] gruCNN & [1, 81, 128, 256] \\ 
 6& [$2\times1$] Maxpool&[$2\times1$] Maxpool&[$2\times1$] Maxpool& [1, 41, 128, 256]\\
 7 & [$3\times3$] CNN & [$3\times3$] CNN& [$3\times3$] gruCNN & [1, 41, 128, 256]\\ 
 8 & [$3\times3$] CNN & [$3\times3$] CNN& [$3\times3$] gruCNN & [1, 41, 128, 256]\\ 
 9 &  FC &  LSTM + FC&  FC & [1, 161, 128, 1]\\
 \bottomrule
\end{tabular}}}
\end{table}

\textbf{Data Set (Training and Testing):} The speech set is a selection of ten British English speakers -- both male and female -- from the Voice Bank speech corpus~\cite{veaux2013voice}, each of which has around 400 clean utterances. Eight speaker's data were used for training, and the remaining two (one male and one female) were reserved for performance testing. The noisy mixtures were created manually. The noises are from~\cite{loizou2013speech}, which contains 20 different types of common environmental noises. Fourteen of which were used for the training, and the remaining six were used as the unseen noises, under which the models are tested. For training set mixtures, each speech sample was masked by a random training set noise at a random SNR point from [0, 5, 10, 15, 20] dB. A similar process has been followed for the test set, but with the unseen noises at unseen SNR points of [2.5, 12.5, 22.5] dB.    

Although the original speech were sampled at 48kHz, it was down-sampled to 16 kHz for our experiment as in~\cite{muhammed2019non}\cite{rethage2018wavenet}. The 16kHz sampled signals were framed into 20 ms frames with 10 ms overlap. The frames were Fourier transformed into 320 points. The log-power spectra feature is the domain on which the enhancement task is modeled~\cite{portnoff1980time}. Therefore, the frequency dimension of input spectrum is halved to 161 points, due to the spectral symmetry.

\textbf{Model Training}:
All the comparing models are trained in an end-to-end mode, where the losses are computed directly between the magnitudes of predicted ($\hat{Y}(k,t)$) and target ($Y(k,t)$) STFT components. For each noisy-clean training set pair $(X, Y)$, the model parameters are optimized by minimizing the mean square error (MSE) objective function.

\begin {equation}
L_{X,Y}=\frac{1}{T \times K}\sum_{t=1,f=1} ^{t=T,f=K}(|Y(k,t)|-|\hat{Y}(k,t)|)^{2}
\label{eq6}
\end{equation}
where K denotes the dimension of frequency axis that is 161, and the variable T is the number of time frames recurrently generated in the training process; which has been set to T = 128. The T value for testing varies based on the input signal duration for the recurrency is being modeled over the temporal axis. The loss was minimized by the Adam optimizer~\cite{kingma2015adam} with an exponentially decaying learning rate method with learning rate = 0.001, decay steps = 20,000 and decay rate = 0.99. 

For objective evaluation of processed samples, the perceptual evaluation of speech quality (PESQ) metric (ITU P.862.2)~\cite{hu2007evaluation} that measures the quality, and the short-time objective intelligibility (STOI)~\cite{taal2011algorithm} that measures the intelligibility, are considered. The composite quality of the model predictions (COVL) has also been measured~\cite{loizou2013speech}, which reports a compound count of the noise reduction and speech restoration. Besides, the SNR intelligibility gain through model processing is measured by the segmental SNR (SSNR) score~\cite{loizou2013speech}. Subjectively, the quality of enhanced samples were measured by the mean opinion score (MOS). In total, 20 participants (non-native English speakers) listened to and assigned the individual perceptual score based on the noise artifacts present, in a scale of 0 -- 5 (0 -- very annoying artifacts , 5 -- no artifacts at all).  

\begin{figure*}
\captionsetup[subfigure]{labelformat=empty}
  \centering
    \begin{subfigure}{\textwidth}
        \includegraphics[width=\linewidth]{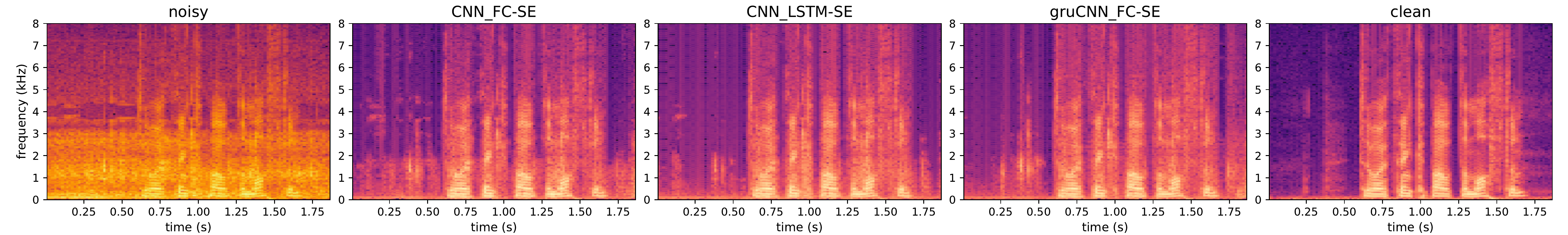}
        \caption{}
        \label{fig:a}
    \end{subfigure}
    \newline
    \vspace{-.4cm}
    \begin{subfigure}{\textwidth}
        \includegraphics[width=\linewidth]{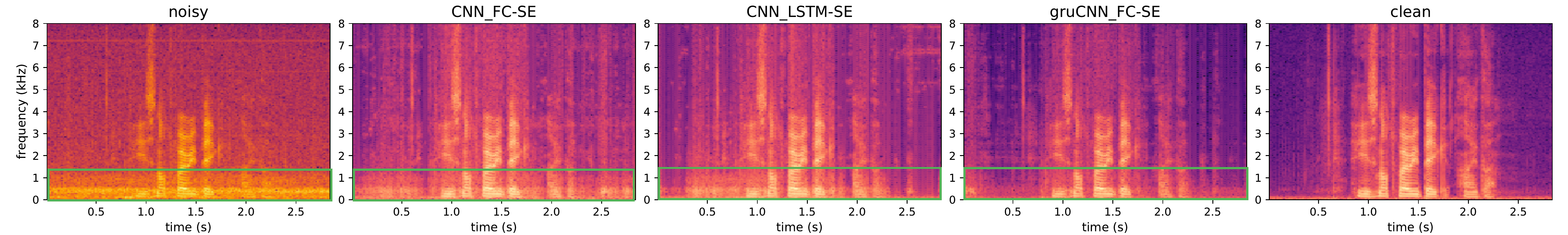}
        \caption{}
        \label{fig:b}
    \end{subfigure}
    \vspace{-.4cm}
    \caption{Model enhancement under construction (upper panel) and street (lower panel) noise}
    \label{fig.2}
\end{figure*}

\section{Results and Discussion}
\label{Sec.4}
The objective scores averaged over the test samples at each noise condition are displayed in TABLE~\ref{tab:Table-2}. Along with the processing types, the scores of unprocessed noisy speech are also provided to better understand the relative processing gain. Compared to CNN\_FC--SE architecture, which does not incorporate any form of recurrency described in~(\ref{eqn3}), the suggested gruCNN\_FC--SE model with recurrency modelled in the feature extraction layers has distinctly outperformed on all the metrics. This gain is almost consistent across the noise conditions. 
With the inclusion of global recurrency by CNN\_LSTM--SE, the performance has improved over CNN\_FC--SE. This broadly conveys the benefits that can be achieved through temporal inclusive modeling of speech.

\begin{center}
\begin{table}[]
\caption{Objective measures for unseen noise conditions}
 \label{tab:Table-2}
 \centering{
 \ra{1.55}
\scalebox{0.8}{
\begin{tabular}{c c c c c c} 
     \toprule
{\bf Noise level}& {\bf Metric} & {\bf Noisy} & {\bf CNN\_FC-SE} & {\bf CNN\_LSTM-SE} & {\bf gruCNN\_FC-SE} \\ 
      \toprule
\multirow{4}{4em}{{\bf 2.5 dB}}  &PESQ &1.20 &1.41 &1.51  &{\bf 1.57} \\
	&STOI &0.68 &0.71 &0.72  &{\bf 0.74}  \\ 
	& COVL &1.58 &1.96 &2.15  &{\bf2.22}   \\ 
	& SSNR & - 3.63 &2.39 &3.20  &{\bf 3.94}   \\ 
\hline
\multirow{4}{4em}{{\bf 12.5 dB}}  &PESQ &1.49 &1.87 &2.01  &{\bf 2.08} \\
	&STOI &0.77 &0.78 &0.79  &{\bf 0.80}        \\ 
	& COVL &2.11 &2.59 &2.74  &{\bf2.83}   \\ 
	& SSNR &3.24 &7.61 &7.85  &{\bf  8.96}   \\
\hline
\multirow{4}{4em}{{\bf 22.5 dB}}  &PESQ &2.27 &2.47 &2.58  &{\bf 2.66} \\
	&STOI &0.85 &0.83 &0.84 &{\bf 0.85}        \\ 
	& COVL &3.05 &3.20 &3.30  &{\bf 3.41}   \\ 
	& SSNR & 12.26 &11.21 &11.14  &{\bf12.83}   \\ 
 \bottomrule
\end{tabular}}}
\vspace{-5mm}
\end{table}
\end{center}
\vspace{-0.8cm}

When comparing the two recurrent models, the proposed gruCNN\_FC--SE, that is concerned of the bin-wise recurreny factor, has shown better enhancement over CNN\_LSTM--SE. Even at the higher SNR point of 22.5dB, where the noise attributes are expected to be mild, gruCNN\_FC--SE model elicited noticeable enhancement, showing an SSNR intelligibility gain of up to 1.5 dB over the other methods. This gain must be attributed to the qualitative restoration of speech components with the suggested feature extraction strategy.

Regarding the consistency of model predictions in different noise types, the model enhancements under the two unseen noise conditions are plotted in Fig.~\ref{fig.2}. The upper panel shows construction noise (type--1) while the lower panel refers to street noise (type--2). 
Since type--1 noise is quite stationary and has the spectral energy that is distributed uniformly in a very wide frequency band (0 - 3 kHz),
it is straightforward for a network to get a frequency smoothed estimate of the noise statistics. While type--2 noise (street) are highly localized at the lower band (0 - 0.5 kHz) of the spectrum (marked by a straight line in Fig.~\ref{fig.2}).
Unless the model looks into the local statistics of the spectrum, these noise activities could easily be miss-classified as speech events. We suggest that this explains the performance of CNN\_FC--SE and CNN\_LSTM--SE, whereas gruCNN\_FC--SE seems to be successful in disentangling out the noise activities by exploiting the local patterns.

The subjective scores of different models are displayed in TABLE~\ref{tab:Table-3}. In line with the objective measures, the suggested gruCNN\_FC--SE model is ranked closer to the clean speech with a score of 3.16 on the five point scale, while there was not any statistically observable difference between the scores of the other two methods.

Pragmatically, the performance gain of a neural model could be argued by the additional parameters that is floated into the modeling. To this end, the parameter counts of different models are shown in TABLE~\ref{tab:Table-4}. 
CNN\_FC--SE is the least complex among the models and indeed its performance has been much lower than the other two models. On the other hand, the suggested gruCNN\_FC--SE produces far better enhancement with only 75\% parameters of the CNN\_LSTM--SE.
This reduction in complexity is from the replacement of the fully-connected LSTM cell with the fixed kernels of gruCNN to model the temporal flow.
All of which indicates the potentiality to have it implemented on computationally constraint applications, like hearing aid. A Tensorflow implementation and enhanced samples from the model are provided at \footnote{\url{https://www.csd.uoc.gr/~shifaspv/IEEE_Letter-demo}} \footnote{\url{https://github.com/shifaspv/gruCNN-speech-enhancement-tensorflow}}.

\begin{table}[th]
\caption{Mean opinion score (MOS) with standard error}
 \label{tab:Table-3}
 \centering{
 \ra{1.6}
\scalebox{0.77}{
\begin{tabular}{c c c c c c} 
     \toprule
 {\bf Metric} & {\bf Noisy}& {\bf CNN\_FC-SE} & {\bf CNN\_LSTM-SE} & {\bf gruCNN\_FC-SE} & {\bf Clean}\\ 
      \toprule
 MOS  & 2.01$\pm$0.97 & 2.75$\pm$0.92  & 2.77$\pm$0.89  & \bf {3.16$\pm$0.92} &  {4.86$\pm$0.42} \\
  \bottomrule
\end{tabular}}}
\vspace{-2mm}
\end{table}

\begin{table}[th]
\caption{Model parameters count in Million (M)}
 \label{tab:Table-4}
 \centering{
 \ra{1.3}
\scalebox{0.95}{
\begin{tabular}{c c c c} 
     \toprule
 {\bf Metric} & {\bf CNN\_FC-SE} & {\bf CNN\_LSTM-SE} & {\bf gruCNN\_FC-SE}\\ 
      \toprule
 Parameters  &11.13M &36.10M  &{ 27.22M} \\
  \bottomrule
\end{tabular}}}
\end{table}
\vspace{-0.2cm}

\vspace{-0.3cm}
\section{Conclusion}
\label{conclusion}
In this letter, we presented a new concept of recurrent feature extraction that is found to be beneficial for single-channel speech enhancement. In contrast to the traditional independent modelling of feature extraction and temporal recurrency, the suggested module with recurrent connections inside is proved to be robust to unseen noise conditions and efficient. The subjective and objective evaluations have confirmed the benefits that the suggested recurrent feature extraction technique elicited. While at the same time, the parameter complexity of the suggested model is reduced by 25\% compared to the traditional model. 


\ifCLASSOPTIONcaptionsoff
  \newpage
\fi


\bibliographystyle{IEEEtran}
\bibliography{mybib}
%
%
%

%






\end{document}